\documentclass[prl,aps,twocolumn,floats,showpacs]{revtex4}%
\usepackage{epsfig,latexsym,amssymb,amsmath,amsbsy,graphics,graphicx}
\usepackage{dcolumn,bm,amsfonts}
\usepackage{amsmath}
\usepackage{amsfonts}
\usepackage{amssymb}
\usepackage{graphicx}%
\begin{document}
\title{Disorder Screening in Strongly Correlated Systems}
\author{D. Tanaskovi\'{c} and V. Dobrosavljevi\'{c}}
\affiliation{Department of Physics and National High Magnetic Field Laboratory \linebreak
Florida State University, Tallahassee, Florida 32306}
\author{E. Abrahams and G. Kotliar}
\affiliation{Center for Materials Theory, Serin Physics Laboratory, Rutgers University, 136
Frelinghuysen Road, Piscataway, New Jersey 08854}

\begin{abstract}
Electron-electron interactions generally reduce the low temperature
resistivity due to the screening of the impurity potential by the electron
gas. In the weak-coupling limit, the magnitude of this screening effect is
determined by the thermodynamic compressibility which is proportional to the
inverse screening length. We show that when strong correlations are present,
although the compressibility is reduced, the screening effect is nevertheless
strongly enhanced. This phenomenon is traced to the same non-perturbative
Kondo-like processes that lead to strong mass enhancements, but which are
absent in weak coupling approaches. We predict metallicity to be strongly
stabilized in an intermediate regime where the interactions and the disorder
are of comparable magnitude.

\end{abstract}
\maketitle

Transport in disordered metals has been studied for many years, and
substantial theoretical and experimental understanding has been achieved
\cite{Lee} in the case of weak disorder and in the regime of weak
electron-electron interaction. Much less is known about situations with strong
electronic correlations; here, most research has concentrated on clean
systems.

Recent experiments on two-dimensional (2D) electron gases in zero magnetic
field \cite{Abrahamsrmp} have led to considerable renewed interest in
electronic systems close to metal-insulator transitions (MITs). In these
systems, well-defined metallic behavior (positive temperature coefficient of
resistivity, $d\rho/dT >0$) has been observed in the low density regime, and
is characterized by a surprisingly large (up to a factor of ten) drop of
resistivity at low temperatures. Here, the electron-electron interactions
represent the largest energy scale in the problem \cite{Abrahamsrmp}; this is
emphasized by recent reports of substantial mass enhancement from several
complementary experiments \cite{mass}.

Some simple microscopic mechanisms that can produce such a resistivity drop
relate to temperature-dependent screening
\cite{DasSarma,Dolgopolov,Herbut,Zala} of the random potential. This effect
obtains even in the simplest Hartree-Fock (HF) treatment of interactions,
which represents the basis for the standard (Lindhard) screening theory, and
applies equally well to both short-range \cite{Herbut} and long-range
\cite{DasSarma} forces. In this picture, the screening length is inversely
proportional to the thermodynamic compressibility of the system, which
therefore controls the magnitude of disorder screening. This observation
immediately brings into question the relevance of such mechanisms in the
regime of strong correlation, in particular close to interaction-driven MITs.
Here, one expects substantial mass enhancements, but at the same time a
significant decrease of the compressibility \cite{Georges}. Then, if
applicable, standard screening theory (as in HF) would predict\textit{ weak
}disorder screening precisely where mass is enhanced - in contrast to
experiments, where the resistivity drop persists in that region (see Fig.\ 3, below).

In this letter, we examine the screening of the impurity potential
by focusing on a model where a reliable and controlled treatment
of strong correlations is available. This is possible within
dynamical mean-field theory (DMFT) \cite{Georges}, which is
formally exact in the limit of large coordination. To investigate
the regime of strong correlation, we examine a disordered Hubbard
model in the vicinity of the Mott transition, a model which has
recently been argued \cite{phillips,Herbut,dolgomott,mitglass} to
provide an appropriate description of interaction effects near the
2D-MIT. Our results demonstrate that: (1) If DMFT equations are
solved within the HF approach, we reproduce the results of
standard screening theory in qualitative and even semiquantitative
detail; in this picture, screening is strongly suppressed close to
the Mott transition. (2) A more accurate solution agrees with HF
results far from the Mott transition, but finds diametrically
opposite results in the regime of strong correlation. Here, while
compressibility is reduced, both the disorder screening and the
effective mass are strongly enhanced. (3) The enhanced screening
strongly stabilizes metallic behavior in the intermediate regime
where the disorder and interactions are of comparable magnitude.

We consider a disordered Hubbard model described by the Hamiltonian
\begin{equation}
H=-\sum_{ij\sigma}t_{ij}c_{i\sigma}^{\dagger}c_{j\sigma}+\sum_{i\sigma
}\varepsilon_{i}n_{i\sigma}+U\sum_{i}n_{i\uparrow}n_{i\downarrow}.
\label{hamiltonian}%
\end{equation}
Here $t_{ij}$ are the hopping matrix elements, $c$ and $c^{\dagger}$ are
fermionic creation and annihilation operators, $n=c^{\dagger}c$ is the number
operator, and $\sigma$ labels the spin projection. $U$ represents the Hubbard
on-site repulsion, and the disorder is introduced by random site energies
$\varepsilon_{i}$, as specified by a distribution function $P(\varepsilon
_{i})$.

\textit{Disorder renormalization (screening).} Within DMFT \cite{Georges}, a
quasiparticle is characterized by a self-energy function $\Sigma(\omega)$,
which is assumed to be purely local (momentum independent). In a random system
\cite{dmftdis}, this quantity though still local, is now site-dependent
$\Sigma_{i}(\omega)=\Sigma(\omega,\varepsilon_{i})$ which explicitly depends
only on the corresponding local site energy $\varepsilon_{i}$ \cite{cavity}.
The renormalized disorder potential (as seen by the quasiparticle at the Fermi
energy) can thus be defined by
\begin{equation}
v_{i}(\varepsilon_{i})=\varepsilon_{i}+\Sigma_{i}(\omega=0)-\mu-\delta\mu,
\end{equation}
where $\mu$ is the chemical potential. These renormalized energies are defined
with respect to a reference energy $\delta\mu$ chosen such that $\overline
{v_{i}}=\int d\varepsilon_{i}\,P(\varepsilon_{i})v_{i}(\varepsilon_{i})=0,$
i.e. that their site-average vanishes. In a case of particle-hole symmetry
$\delta\mu=0$. Since there are no vertex corrections within DMFT, the $T=0$ dc
resistivity \cite{Georges,dmftdis} depends only on the variance of this
renormalized disorder, viz. $\rho\sim\overline{v_{i}^{2}}$. The self-energies
$\Sigma_{i}(\omega)$ must be calculated by solving an ensemble of Anderson
impurity problems supplemented by a self-consistency condition
\cite{Georges,dmftdis}.

\textit{Weak coupling} (\textit{Hartree-Fock}) \textit{solution. }In the HF
approximation for the Hubbard model, $\Sigma_{i}(\omega)=U\left\langle
n_{i}\right\rangle ;$ here $\left\langle \cdots\right\rangle $ represents the
quantum average (for a given disorder configuration). Within DMFT, the local
occupation $\left\langle n_{i}\right\rangle $ depends only on the local
(renormalized) site energy, and for moderate disorder \cite{Herbut} we find
\begin{equation}
\frac{\,\,\overline{v_{i}^{2}}\,\,}{\overline{\varepsilon_{i}^{2}}}=\left[
1+U\chi_{ii}\right]  ^{-2},\label{HF}%
\end{equation}
where $\chi_{ii}=-\partial\left\langle n_{i}\right\rangle /\partial
\varepsilon_{i}$ is the local compressibility.\cite{zimanyi} This result,
although based on a local approximation, proves to provide qualitative and
even semiquantitative agreement with more standard screening theory. Just as
in Refs. \cite{DasSarma,Dolgopolov,Herbut}, for reasonable values of the
interaction strength ($U\lesssim$ bandwidth), the screening of the random
potential cannot be very large (resistivity drop by a factor of two at
most\cite{DasSarma}).
%Even more importantly, the screening depends
%crucially on the (local) compressibility, a quantity that is very small
%in the strongly correlated regime.

\begin{figure}[ptbh]
\begin{center}
\includegraphics[
trim=0.000000in 0.000000in 0.000000in -0.617320in,
height=2.2059in,
width=2.919in
]{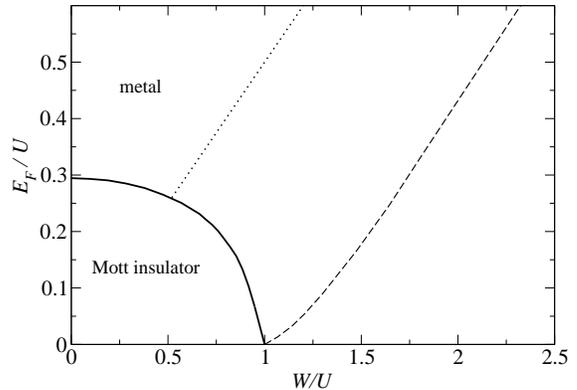}
\end{center}
\caption{DMFT phase diagram of the Hubbard model with random site energies.
The Mott insulator can be suppressed by sufficiently strong (bare) disorder
$W>U.$ Also shown is an estimate of the regime where Anderson localization
effects are important, as obtained by comparing the Fermi energy $E_{F}$ to
bare disorder $W$ (dotted line) or screened disorder $\widetilde{W}$ (dashed
line). Localization is strongly suppressed by correlation effects in the
intermediate regime where the disorder is comparable to the on-site repulsion
$U.$}%
\end{figure}

\textit{Strongly correlated regime.} We approach a $T=0$ Mott transition at
half-filling. In general, the DMFT equations cannot be solved in closed form,
but extensive work \cite{Georges} has shown that most qualitative and even
quantitative features of the $T=0$ solution can be reproduced using simple
analytical approximations. Here, we use the four-boson mean-field method of
Kotliar and Ruckenstein (KR),\cite{K-R} which is equivalent to the well-known
Gutzwiller variational approximation, but can be readily generalized to
disordered systems. This approach provides a parameterization of the
low-energy (quasiparticle) part of the local Green function which, in our
case, takes the form $(\hbar= c =1)$
\begin{equation}
G_{i}(\omega_{n})=\frac{Z_{i}}{i\omega_{n}-\widetilde{\varepsilon}_{i}%
-Z_{i}\Delta(\omega_{n})}.
\end{equation}
Here, the local quasiparticle weight $Z_{i}=2\left[  1-(e_{i}^{2}-d_{i}%
^{2})^{2}\right]  ^{-1}(e_{i}+d_{i})^{2}\left[  1-(e_{i}^{2}+d_{i}^{2})\right]
$, as well as parameters $e_{i}$, $d_{i}$ and $\widetilde{\varepsilon_{i}}$
are site-dependent \cite{zimanyi} quantities, determined by the KR equations
\cite{K-R}
\begin{align}
\!\!\!\!-\frac{\partial Z_{i}}{\partial e_{i}}\frac{1}{\beta}\sum_{\omega_{n}%
}\Delta(\omega_{n})G_{i}(\omega_{n})  &  =Z_{i}\left(  \mu+\widetilde
{\varepsilon}_{i}-\varepsilon_{i}\right)  e_{i},\label{lambda}\\
\!\!\!\!-\frac{\partial Z_{i}}{\partial d_{i}}\frac{1}{\beta}\sum_{\omega_{n}%
}\Delta(\omega_{n})G_{i}(\omega_{n})  &  =Z_{i}\left(  U\!-\!\mu
\!-\!\widetilde{\varepsilon}_{i}+\!\varepsilon_{i}\right)  d_{i}%
,\,\,\,\,\,\label{d}\\
\frac{1}{\beta}\sum_{\omega_{n}}G_{i}(\omega_{n})  &  =\frac{1}{2}%
Z_{i}(1-e_{i}^{2}+d_{i}^{2}). \label{sumrule}%
\end{align}
Half-filling can be enforced by the requirement $\overline{e_{i}^{2}%
}=\overline{d_{i}^{2}}.$ Finally, a self-consistency condition \cite{dmftdis}
determines the ``hybridization function"  $\Delta(\omega)=\Delta_{o}\left(
\omega-\Sigma_{av}(\omega)\right)  $  describing the environment of a given
site.  Here $\Delta_{o}(\omega)=$ $\omega+\mu-\left[  G_{o}(\omega)\right]
^{-1}$, $G_{o}(\omega)$ being the lattice Green's function corresponding to no
disorder and $U=0.$ The ``average" self-energy $\Sigma_{av}(\omega)$ is
defined via the disorder-averaged Green's function $\overline{G}(\omega)=\int
d\varepsilon_{i}\,P(\varepsilon_{i})G_{i}(\omega)$: $\overline{G}%
(\omega)\equiv G_{o}\left(  \omega-\Sigma_{av}(\omega)\right) .$ Thus
$\Sigma_{av}(\omega)=\omega+\mu-\Delta(\omega)-\left[  \overline{G}%
(\omega)\right]  ^{-1}$. It follows that $\delta\mu=\int d\varepsilon
_{i}\,P(\varepsilon_{i})\!\,\widetilde{\varepsilon}_{i}/Z_{i},$ and
$v_{i}=\!\widetilde{\varepsilon}_{i}/Z_{i\,}-\delta\mu.$

\textit{Phase diagram.} These equations can be easily solved for an arbitrary
band structure and distribution of disorder, but such details do not affect
the qualitative form of the solution, which proves to depend only on the
presence or absence of particle-hole symmetry. We first examine the
particle-hole symmetric situation, and as an illustration we concentrate on a
model with a semi-circular band of width $4t$ at half-filling, and site
energies uniformly distributed in the interval $(-W/2,W/2)$. The resulting
$T=0$ phase diagram, as obtained from a full numerical solution, is shown in
Fig. 1. The Mott insulating phase is completely suppressed for $W>U,$ since
the disorder tends to fill in the Hubbard-Mott gap. The phase boundary
separating the correlated metal and the Mott insulator (full line) is
identified by the simultaneous vanishing of the quasiparticle weights $Z_{i}$
on all lattice sites. All quantities display simple critical behavior close to
this phase boundary, the form of which can be analytically obtained for weak
disorder, but which proves to remain qualitatively correct along the entire
critical line. To second order in $\varepsilon_{i}$, we find $Z_{i}%
=Z_{0}(1+C\varepsilon_{i}^{2})$, where $Z_{0}\equiv Z_{i}(\varepsilon
_{i}=0)=2(1-U/U_{c})$ linearly goes to zero at the critical interaction
$U_{c}(W)=U_{c}^{0}(1+\frac{2}{5}C\overline{\varepsilon_{i}^{2}})$. This
corresponds to the well-known effective mass enhancement ($m^{\ast}\sim1/Z)$
in the strongly correlated regime. Here the constant $C=10/(U_{c}^{0})^{2}$,
and $U_{c}^{0}=64t/3\pi$ for the considered band structure.

\textit{Scattering rate.} To determine the effects of correlations on
transport, we calculate the \mbox{scattering rate $1/\tau=$}
$-2\,\mbox{Im}\Sigma_{av} (\omega=0)=-2\,\mbox{Im}G_{o}(0)\overline{v_{i}^{2}%
}$. We find that scattering is strongly reduced near the Mott transition,
corresponding to correlation-enhanced screening. To show this analytically, we
note that in the critical regime the spectral weight corresponding to
$\Delta(\omega)$ is of the order of $Z_{o}.$ We expand Eq.~(\ref{sumrule})
with respect to $\widetilde{\varepsilon}_{i}/Z_{i}$: To leading order,
$\tilde{\varepsilon}_{i}/Z_{i}\sim-e_{i}^{2}+d_{i}^{2}\sim\mathcal{O}(Z_{i})$.
Therefore $\widetilde{\varepsilon}_{i}/Z_{i}\equiv v_{i}\sim Z_{i}%
\rightarrow0$ as $U\rightarrow U_{c}(W)$, and we conclude that random site
energies are \textit{perfectly screened} at the metallic side of the Mott
transition \cite{realistic}. The same conclusion remains valid even for an
asymmetric distribution of disorder, in which case particle-hole symmetry is
restored as the transition is approached. For weak disorder we obtained a
simple formula for the scattering rate close to the Mott transition
\begin{equation}
\frac{1}{\tau}=\frac{2}{t}\left[  \frac{3\pi t}{4U_{c}}\left(  1-\frac
{U}{U_{c}}\right)  \right]  ^{2}\overline{\varepsilon_{i}^{2}}.
\label{sircular}%
\end{equation}
Typical numerical results obtained for disorder ranging from weak to strong
are shown in Fig.~2(a). We find that disorder is strongly screened even
relatively far from the transition, for $U/U_{c}(W)\gtrsim0.5.$

\begin{figure}[t]
%[ptbh]
\par
\begin{center}
\includegraphics[
height=4.2in,
width=2.919in
]{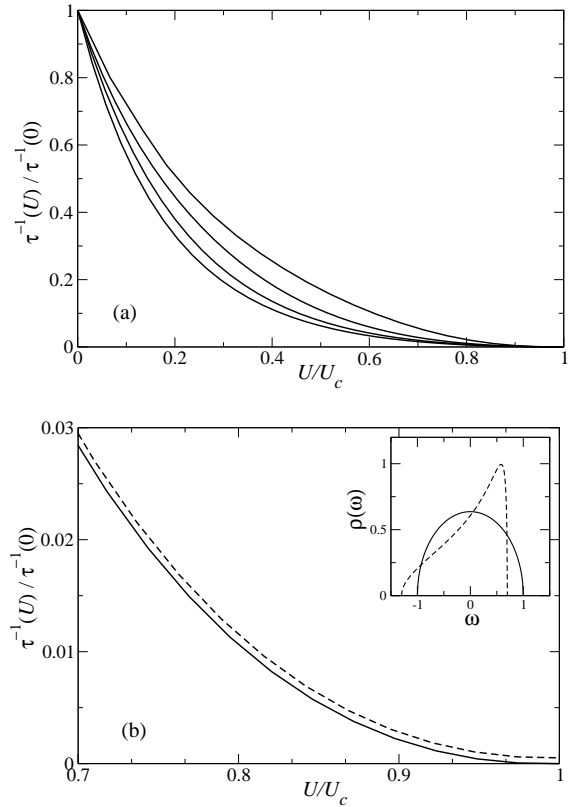}
\end{center}
\caption{Scattering rate normalized with the noninteracting value. (a) From
the lower to the upper curves: $W=0.1,1,2,4$. (b) Results close to $U_{c}$ for
the particle-hole symmetric lattice (full line), and the asymmetric lattice
(dashed line), with $W=1$. The inset shows the density of states in these two
cases.}%
\end{figure}

\textit{Particle-hole asymmetry. }For a particle-hole asymmetric model
(generally appropriate for realistic materials), we find that for
$U\rightarrow U_{c}(W)$ the energies $v_{i}$ remain finite, although still
much smaller than $W$. The disorder screening nevertheless remains very
strong, which persists even in the case of fairly strong particle-hole
asymmetry and strong disorder. Typical numerical results are shown in Fig.
2(b), where the full line corresponds to the symmetric semicircular density of
states, and the dashed line is obtained for the particle-hole asymmetric
lattice described by the Green function $G_{o}(\omega)=\left[  \omega
-t^{2}G_{o}(\omega+a)\right]  ^{-1}$. In this plot, the asymmetry parameter
$a$ is chosen to be $0.15D$ (strongly asymmetric lattice), where $D$ is the
bandwidth. For weak disorder and moderate particle-hole asymmetry we were able
to obtain a quantitative estimate of the screening effect, and we find that
close to the transition the scattering rate approaches a very small asymptotic
value $1/\tau\sim(\delta\mu)^{2}\overline{\varepsilon_{i}^{4}}\sim W^{4}$,
instead of $ 1/\tau\sim W^{2}$ as for weak interactions. In addition, for any
realistic lattice at half filling, $\delta\mu$ is a small number ($\leq0.1$),
explaining the smallness of the scattering rate. \begin{figure}[t]
%[ptbh]
\par
\begin{center}
\includegraphics[
height=2.0232in,
width=2.919in
]{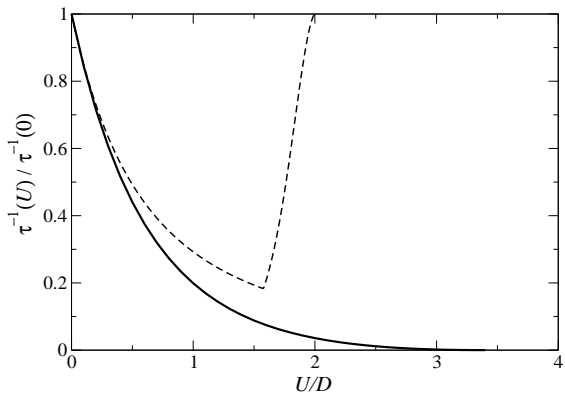}
\end{center}
\caption{Normalized scattering rate as a function of $U$, from the full DMFT
solution (full line), and the corresponding Hartree-Fock (HF) approximation
(dashed line). For moderate interaction both methods predict the same
screening, but diametrically opposite results are obtained in the
strongly-correlated regime, where DMFT predicts enhanced screening, while a
strong suppression is obtained within HF theory. }%
\end{figure}

\textit{Breakdown of conventional theory. }In the strongly-correlated regime
our DMFT results are in sharp discrepancy with results obtained within
Hartree-Fock, as shown in Fig.~3. For small $U$, both methods give similar
results, but closer to the transition HF theory (dashed line) predicts a
reduced disorder screening, while full DMFT (full line) shows that the
screening remains strongly enhanced. The reduction of screening found in HF
reflects the decrease of the compressibility near the Mott transition. This is
a result of a Stoner instability in the magnetic HF solution, which sets in
for $U=1/\rho_{0}(0)=\pi D/2$ as a precursor to a gap opening at the
transition \cite{stoner}. We emphasize that the direct relation between
compressibility and screening as in Eq.~$\left(  \ref{HF}\right) $ is a
general feature of weak-coupling approaches, and its applicability is
seriously limited in the strongly-correlated regime.

\textit{Enhanced screening as ``Kondo pinning". }In the DMFT approach
\cite{Georges} that we use, the solution of the full Hubbard model is mapped
to solving an ensemble \cite{dmftdis} of auxiliary Kondo-Anderson impurity
problems. Accordingly, the approach to the Mott transition can be described as
the decrease of the local Kondo temperature, corresponding to the reduction of
the local quasiparticle weight. In this language, the renormalized energy
level $v_{i}$ can be identified as the position of the Kondo resonance, which
is well-known to ``pin" to the Fermi energy in the Kondo limit $Z_{i}%
\rightarrow0.$ We can thus interpret the surprising enhancement of disorder
screening in the strongly-correlated regime as reflecting the non-perturbative
Kondo physics captured by our DMFT method, but not by standard weak-coupling
theories. This mechanism is very closely related to the Kondo-enhancement of
resonant tunneling through quantum dots, as observed in recent experiments
\cite{quantumdots}. Our discussion makes it clear why site randomness is
strongly suppressed, but also indicates that if additional hopping randomness
\cite{dmftdis} is introduced, the same mechanism would \textit{not} apply, as
we have verified by explicit calculations. In realistic systems we expect the
disorder to be strongly but not perfectly screened even in the vicinity of the
Mott transition.

\textit{Screening and localization. }The DMFT approach is too simple to
describe Anderson localization effects, which cannot be neglected for strong
enough randomness. Nevertheless, it is interesting to estimate the disorder
strength necessary for localization. In the absence of interactions,
localization is expected \cite{Lee} to set in when the disorder scale $W$ is
comparable to the kinetic (i.e. Fermi) energy, as indicated by a dotted line
in Fig.~1. However, we have shown that correlations lead to strong screening,
with a renormalized disorder scale $\widetilde{W} \sim(\, \overline{v_{i}^{2}%
})\,^{1/2}\ll W,$ which we can numerically compute for any $U$ and $W$, and
analytically in several limits. In particular, in the atomic limit
($E_{F}\rightarrow0$), we find $\widetilde{W}\sim\left(  1-U/W\right)  ^{3/2}%
$. In the presence of interactions, the onset of localization should be
estimated by comparing $\widetilde{W}$ to $E_{F}$, and the resulting
\ boundary is shown by a dashed line in Fig. 1. Interestingly, the metallic
phase is found to be strongly stabilized by screening in the intermediate
regime $W\sim U.$ Of course, such interplay of correlation and localization
should be studied in more detail by extensions of DMFT which can explicitly
incorporate the localization effects \cite{mottand}, but this fascinating
issue remains a challenge for future work.

We thank A. Georges, S. Das Sarma, D. Popovi\'{c}, and Z.
Radovi\'c for useful discussions. This work was supported by the
National High Magnetic Field Laboratory and NSF grants DMR-9974311
and DMR-0234215 (DT,VD), DMR-9976665 (EA), and DMR-0096462 (GK).
VD and EA\ also thank the Aspen Center for Physics where part of
this work was carried out.


\begin{thebibliography}{99}
\bibitem {cavity}Within DMFT for disordered systems \cite{dmftdis}, the
environment of a given site is replaced by its disorder-averaged value.
\bibitem {realistic}We emphasize that the approach to the Mott transition
considered in this work does not correspond to the physical trajectory
describing the reduction of density in 2D electronic systems
\cite{Abrahamsrmp}. There ($E_{F}/U)\sim n^{1/2},$ while $(W/U)\sim n^{-1/2},$
where $n$ is the electron density. Applications of our theory to experimental
situations will be addressed elsewhere.
\bibitem {stoner}This ferromagnetic instability is well-known to be an
artifact of the HF theory for a half-filled Hubbard model, and follows from
the neglect of the correlation energy in this approach.
\end{thebibliography}
\end{document}